\documentclass[reprint,
 amsmath,amssymb,
 prl,
]{revtex4-1}
\usepackage{ mathrsfs }
\usepackage{graphicx}% Include figure files
\usepackage{dcolumn}% Align table columns on decimal point
\usepackage{bm}% bold math
\usepackage{xcolor}
\usepackage{physics} 
\usepackage{hyperref}
\usepackage{cleveref}

\usepackage{bm}

\newcommand{\lag}{\mathscr{L}}

\definecolor{agmc}{rgb}{0.38, 0.31, 0.86}

\begin{document}

%\preprint{APS/123-QED}

\title{A new consistent Neutron Star Equation of State from a Generalized Skyrme model}% Force line breaks with \\
\author{Christoph Adam}
\author{Alberto Garc\'ia Mart\'in-Caro}
\author{Miguel Huidobro}%
\author{Ricardo V\'azquez}
\affiliation{%
Departamento de F\'isica de Part\'iculas, Universidad de Santiago de Compostela and Instituto
Galego de F\'isica de Altas Enerxias (IGFAE) E-15782 Santiago de Compostela, Spain
}%
\author{Andrzej Wereszczynski}
\affiliation{
Institute of Physics, Jagiellonian University, Lojasiewicza 11, Krak\'ow, Poland
}%
%\author{Delta Author}\affiliation{%Authors' institution and/or address\\This line break forced with \textbackslash\textbackslash
%}
\date{\today}% It is always \today, today,
             %  but any date may be explicitly specified

\begin{abstract}
We propose a new equation of state for nuclear matter based on a generalized Skyrme model which is consistent with all current constraints on the observed properties of neutron stars. This generalized model depends only on two free parameters related to the ranges of pressure values at which different submodels are dominant, and which can be adjusted so that mass-radius and deformability constraints from astrophysical and gravitational wave measurements can be  met. Our results support the  Skyrme model and its generalizations as good candidates for a low energy effective field-theoretic description of nuclear matter even at extreme conditions such as those inside neutron stars.
\end{abstract}

%\keywords{Suggested keywords}%Use showkeys class option if keyword
                              %display desired
\maketitle

%\tableofcontents

\section{Introduction}

The modern understanding of strong interactions in the Standard Model of particle physics is based on the theory of Quantum Chromodynamics (QCD), a non-abelian gauge theory where the fundamental degrees of freedom are carried by the quark and gluon fields. Despite its great success at very high energies, we are unable to achieve the same precision in the low-energy regime using full QCD, since the theory becomes nonperturbative. 
In particular, theoretical computations of the properties of baryons and nuclei from QCD are extremely difficult even for the smallest nuclei, and  phenomenological models are usually employed, instead.

The Skyrme model \cite{Skyrme61} offers an alternative approach to this problem. It constitutes
 a nonlinear field theory of mesons which corresponds to an effective field theory for low-energy QCD in the large $N_c$ expansion.
In the Skyrme model, baryons and nuclei emerge as topological solitons, i.e., classical solutions with localized energy density which are stabilized due to the nontrivial topology of the vacuum manifold. 
As a consequence, many non-perturbative features of low-energy QCD, like the conservation of baryon number, the extended character of nucleons, or the global symmetries of QCD and the corresponding symmetry breaking patterns, follow from built-in properties of the Skyrme model.
This field of research has experienced significant progress in recent years, as different generalizations of the model, like the addition of higher derivative terms \cite{JACKSON1985101} or additional degrees of freedom (DoF)---e.g., vector mesons \cite{ADKINS1984251,MeissnerZahed,Meissner:1987ge,Vecmes2,Sutcliffe:2010et,Ma:2019ery}---, or more general potential terms \cite{Marleau:1990nh}, have been proposed  to better reproduce the observed nuclear properties \cite{Adam_2010,Adam:2013wya,Gillard:2015eia,Gudnason:2016mms,Naya_2018}. Further, improved quantization methods which go beyond the moduli space quantization of spin and isospin have significantly contributed to this recent progress, see, e.g.,  \cite{Halcrow:2016spb,Gudnason:2018ysx,Halcrow:2020gbm}.

On the other hand, the first observations of gravitational waves by LIGO opened a new window for the exploration of matter at ultra high densities, like at the cores of Neutron Stars (NS), which are thought to be the most dense objects allowed by General Relativity (GR) before collapsing to a black hole. Indeed, recent \cite{Abbott_2017}---and prospect---observations of mergers of NS binaries will allow us to constrain the equation of state (EoS) of nuclear matter at such high densities. In particular, since the Skyrme model and its generalizations allow to find star-like solutions when coupled to GR, these observations may serve us to determine whether the (generalized) Skyrme model is a consistent way to describe the properties of nuclei and nuclear matter at a large range of scales in a unified manner. 

Different models for NS as Skyrme solitons have been previously proposed, for example, in \cite{NelmesPiette,Adam_2015a}. These models are interesting from a theoretical point of view, because they allow to obtain the EoS of NS cores from a relatively simple field theoretic description. However, none of the Skyrmion star models present in the literature have achieved a good agreement with current observational data of NS \cite{Naya_2019}. 
In this paper, we present an EoS for NS based on a generalized Skyrme model 
which satisfies {\em all} recent observational constraints of NS, such as the maximum mass limit or the deformability as measured in coalescent binary systems.

In this article we will use units in which $c =1$. For masses and lengths we use either nuclear physics units (MeV and fm) or astrophysical units (solar masses $M_\odot$ and km).

\section{Skyrme crystals}

The Skyrme model is an effective field theory of strong interactions at low energies which emerges in the large $N_c$ limit of QCD. It is defined via the Lagrangian
\begin{equation}
    \lag_{SK}=%\lag_{2}+\lag_{4}+\lag_{0}=
    \frac{-f_\pi^2}{4}\Tr{L_\mu L^\mu}+\frac{1}{32e^2}\Tr{[L_\mu,L_\nu][L^\mu,L^\nu]}-\mu^2\mathcal{U},
    \label{Lsk}
\end{equation}
with $f_\pi$ the pion decay constant and $e$ the Skyrme coupling constant. Also, $L_\mu=U^\dagger\partial_\mu U$ is the left invariant Maurer-Cartan form associated to the SU(2)-valued Skyrme field $U(x)$, and $\mathcal{U}=\mathcal{U}(U)$ is a potential.  For the pion mass potential $\mathcal{U}_\pi = (1/2)\, {\rm tr}\, (1-U)$, the parameter $\mu$ is related to the pion mass $m_\pi$ via $\mu = (1/2) f_\pi m_\pi$.

In order to obtain finite energy configurations, one imposes constant boundary values of $U$ at $\abs{x}\rightarrow \infty$, so that the physically relevant Skyrme field configurations define maps
$
    U:S^3\rightarrow SU(2)\simeq S^3,
$
and thus the Skyrme model presents topological solitons (Skyrmions), whose topological charge equals the topological degree of these maps,
\begin{equation}
    \mathcal{B}=\int B^0 d^3 x, \,
    \text{with}\;
B^\mu=\frac{1}{24\pi^2}\varepsilon^{\mu\nu\rho\sigma}\Tr{L_\nu L_\rho L_\sigma}
\end{equation}
the baryon density current.
The Skyrme model \eqref{Lsk} describes an interacting theory for the Goldstone bosons associated to the (broken) chiral symmetry, but baryons emerge as topological solitons, whose topological charge corresponds to the baryon number \cite{Witten:1983tx}. Furthermore, the Skyrme model has been applied to the study of matter at extremely high densities, required to describe the EoS of NS. To do so, one needs to find the lowest energy solutions of the Skyrme model for the very large baryon number of NS, typically  $N\sim N_{\odot}\sim10^{57}$.

It is well known \cite{Kugler:1989uc,Naya_2019} that the lowest energy solutions of the standard Skyrme model (described by the Lagrangian density \eqref{Lsk}) for very large baryon number consist of crystalline cubic  lattices of $B=4$ Skyrmions---which can be thought of as $\alpha$ particles.
The energy per baryon of such solutions as a function of the lattice parameter of the unit cell, $l$, is given by :
\begin{equation}
    E(l)=E_0\qty[a\qty(\frac{l}{l_0}+\frac{l_0}{l})+b]
    \label{energy/baryon}
\end{equation}
being $a=0.474$ and $b=0.0515$ adimensional quantities whose numerical values were obtained in \cite{CASTILLEJO1989801}. 
We fit the values of energy (per baryon) and lattice length corresponding to the minimum energy configuration, $E_0 = 923.32$ MeV and $l_0^{-3} = n_0 = 0.16$ $\text{fm}^{-3}$, to reproduce the energy per baryon of infinite nuclear matter at nuclear saturation density $n_0$, \cite{Vinhas_2015}. Note that our values  slightly differ from those originally proposed in \cite{CASTILLEJO1989801} due to the different fit \footnote{In \cite{CASTILLEJO1989801} $E_0$ and $l_0$ are fitted to the nucleon in the standard Skyrme model parametrization which, on its part, uses the fit to the nucleon and Delta resonance masses. For our purposes, a fit to infinite nuclear matter is much more natural. In addition, using the (nonrelativistic) rigid rotor quantization to calculate the (highly relativistic) Delta mass is intrinsically problematic \cite{Adam:2016drk}.} \footnote{In principle, the pion mass term $(1/4) m_\pi^2 f_\pi^2 l^3$ should be added, but it turns out that its contribution to the Skyrme crystal is negligible for $l\le l_0$ \cite{NelmesPiette, CASTILLEJO1989801}}. From this expression one may obtain the energy per baryon as a function of the pressure \cite{NelmesPiette}, i.e., the EoS of the Skyrme crystal (at zero temperature).

Indeed, by the thermodynamical definition of pressure at zero temperature,
\begin{equation}
    p=-\pdv{E}{V}\equiv-\pdv{E(l)}{l^3}=-\frac{1}{3l^2}\dv{E(l)}{l},
\end{equation}
we have
\begin{equation}
    p(l)=a\frac{E_0}{3l^2}\qty(\frac{l_0}{l^2}-\frac{1}{l_0}) .
    \label{p(l)} 
\end{equation}
This expression for the pressure vanishes at the finite length $l=l_0$, which is a well-known property of infinite nuclear matter at saturation density $n_0$. Further we shall argue below that the standard Skyrme crystal should provide the leading contribution to the nuclear EoS close to nuclear saturation. This explains the fit of the Skyrme crystal parameters $l_0$ and $E_0$ to the infinite nuclear matter values. 

The above expression $p(l)$ can be inverted (solved for $l$), 
\begin{equation}
    \frac{l_0^2}{l^2}=\frac{1}{2}\left( 1+\sqrt{1+\frac{12}{E_0a}pl_0^3} \right) ,
    \label{y(p)}
\end{equation}
and we may substitute the resulting $l(p)$ into \eqref{energy/baryon} to obtain the energy per baryon of the crystal as a function of the pressure, i.e. the equation of state for the Skyrme crystal (at zero temperature).

\section{The Generalized Skyrme model}
\label{sec:2}
Since it is an effective theory, the Skyrme model can be extended by adding higher order terms to the Lagrangian. The  only possible Lorentz-invariant extra term with at most second order time derivatives of the Skyrme field is \cite{Adam_2010} 
\begin{equation} \label{sextic}
    \lag_6=-\lambda^2\pi^4B_\mu B^\mu,
\end{equation}
with $\lambda$ a coupling parameter. Thus, the generalized Skyrme Lagrangian reads ${\lag_{SK}^{gen}=\lag_{SK}+\lag_6}$.

Unfortunately, neither large $\mathcal{B}$ solutions for $\lag_{SK}^{gen}$ nor the corresponding EoS have been found, to our knowledge. However, at sufficiently high densities---for instance, those which occur at the core of a neutron star, which can reach several times the nuclear saturation density $n_0$---, the sextic term (\ref{sextic}) provides the most important contribution to the EoS, related to the $\omega$ meson repulsion of nuclear matter \cite{Adam:2015lra}. The sextic term alone defines a barotropic perfect fluid with energy density $\rho_6 = \lambda^2 \pi^4 n^2 =p$ (see below), where $p$ is the pressure and $n$ the baryon number density. The EoS $\rho_6 =p$ is maximally stiff with a speed of sound equal to 1, which explains its dominance at high density. 

$\lag_{SK}^{gen}$ has another interesting submodel which will be relevant for us, the so-called BPS (=Bogomolnyi-Prasad-Sommerfield) Skyrme model
$
    \lag_{BPS}=\lag_6-\mu^2\mathcal{U}(U).
$
This model supports topological soliton configurations saturating a BPS energy bound \cite{Adam_2010}, hence the name of the model.
Minimally coupling this submodel to gravity \cite{Adam_2015b}, we obtain its stress-energy tensor 
which still is of the perfect fluid form, $T_{BPS}^{\mu\nu}=(p+\rho)u^\mu u^\nu-p g^{\mu\nu}$, with the following definitions (here $g := \abs{\det{g_{\rho\sigma}}}$),
\begin{equation}
    u^\mu=\frac{B^\mu}{\sqrt{g_{\rho\sigma}B^\rho B^\sigma}},\qquad p=\lambda^2\pi^4g^{-1}g_{\rho\sigma}B^\rho B^\sigma-\mu^2\mathcal{U}.
    \label{pressuredensitydef}
\end{equation}
and $\rho=p+2\mu^2\mathcal{U}$. Further, the proper baryon number density is $n=u^\mu (g^{-\frac{1}{2}}B_\mu ) = \sqrt{g^{-1}g_{\mu\nu}B^\mu B^\nu}$.
Note that this perfect fluid is, in general, non-barotropic, since the potential term $\mathcal{U}$ introduces a dependence on the Skyrme field in $p$ and $\rho$, such that no simple algebraic relation can be found between them. 
 Nevertheless, one may still perform a mean-field approximation and obtain an effective, barotropic EoS for the BPS Skyrme fluid, which offers the interesting possibility to compare the results obtained within the exact and the mean-field approaches \cite{Adam_2015b}.

In the case of interest here, however, we will introduce a constant effective potential $\mu^2 \mathcal{U} = \rho_0$, which is supposed to take into account the effects of the subleading contributions above a certain threshold value $p_{\rm PT}$ for the pressure, see below. This is equivalent to choosing the theta-term potential of ref. \cite{Adam_2015b} and implies the barotropic EoS 
\begin{equation}
\rho = \rho_6 + \rho_0 = \lambda^2\pi^4 n^2 +    \rho_0 = p + 2\, \rho_0
\label{BPSEoS}
\end{equation} 
already at the full field-theory level.

\section{A generalized equation of state}
\label{subsec:hybrid model}

%As stated before, the BPS Skyrme model may be useful to describe nuclear matter at very high densities compared to the nuclear saturation. On the other hand, it is expected that the quadratic (kinetic) and quartic (Skyrme) terms in $\eqref{Lsk}$ become more important as the density drops to the nuclear saturation density, since it is in this regime where the standard Skyrme model has been shown to accurately predict some properties of nucleons and nuclei {\color{red}cite[]}.

Both the standard Skyrme model and the BPS submodel have been previously used to describe nuclear matter inside NS \cite{Naya_2019}. However, it is clear from these attempts that the true equation of state for Skyrme matter should take into account both models in a unified fashion, because the results from approximating the full model with either of the two submodels deviate from the most recent observational data of NS, and do so in opposite directions. For example, the maximum mass of NS are either too small (for pure skyrmion crystals) or too large (for BPS Skyrmion stars) as compared with the current constraints \cite{Naya_2019}.
As explained, the generalized Skyrme model has not been solved yet for large baryon number. Nevertheless, we may still obtain some information of these high baryon number solutions by scaling arguments of the energy terms for the different submodels of the complete Lagrangian.

Indeed, consider for example the case of the Skyrmion crystal, whose energy per baryon is given by \eqref{energy/baryon}, and let $\sigma\in (0,1]$. A scale transformation of the space coordinates of the form $x\mapsto x/\sigma$ can be understood as a mapping between crystalline solutions, respectively, with lattice size $l$ and $\sigma l$. On the other hand, since the lattice length is a function of the pressure, we conclude that two solutions at different pressures $p$ and $p'$ which have a lattice length of $l(p)$ and $l'(p')$ respectively, are related through a scale transformation $\sigma(p,p')$ such that $l'(p')=\sigma(p,p')l(p)$. In particular, any configuration with lattice length $l(p)$ will be related to the zero pressure crystal (minimum energy configuration) via $l(p)=\sigma(p)l_0$, where $\sigma(p)=\sigma(0,p)$ can be seen as a function relating the pressure of the crystal and the scaling parameter. Furthermore, taking into account \eqref{y(p)}, we find
\begin{equation}
    \sigma (p)=\sqrt{\frac{2}{1+\sqrt{1+\frac{12l_0^3}{E_0a}p}}}.
\end{equation}
This expression has indeed the correct limits of $\sigma(p\rightarrow\infty)\rightarrow 0$ and $\sigma(p=0)=1$ .

This equivalence between pressure and scaling allows us to write the energy per baryon of the Skyrmion crystal at any pressure (i.e. $\sigma\neq 1$) as a simple function of $\sigma =l/l_0$,
$    E(\sigma)=a \, E_0\left( \sigma+\sigma^{-1}\right) + b \, E_0 $.
Obviously, the contribution from the term proportional to $\sigma$ becomes negligible for large pressure, whereas the term proportional to $\sigma^{-1}$ dominates in this regime (${\sigma\ll1}$).

Next, consider the sextic term contribution to the energy (and energy per baryon) of a fluid element $\Omega$ 
\begin{equation}
  \frac{E_6}{\mathcal{B}}=\frac{(\int_\Omega d^3 x \sqrt{g}\, \rho_6) }{(\int_\Omega d^3 x \sqrt{g}\,  n)},
\end{equation}
which transforms as $E_6\mapsto \sigma^{-3}E_6$ under a scaling of spacetime coordinates. This implies that the sextic contribution will dominate the energy per baryon at sufficiently high pressure. Therefore, we may assume that a solution of the complete model will provide an EoS which tends to the EoS of the submodel $\rho_6$ at high pressure, with an asymptotic energy per baryon of $E_6/\mathcal{B}=\rho_6/n = \lambda\pi^2\sqrt{p}$. This is, therefore, the asymptotic behavior of the energy per baryon at high pressure also for the full model.

On the other hand, as the pressure decreases to a certain value (which depends on $\lambda$), $E_6/\mathcal{B}$ becomes of the order of the energy per baryon of the Skyrme crystal, and the BPS approximation to the complete solution will start to fail. 
For even lower $p$, the contribution of $E_6/\mathcal{B}$ will be subleading in comparison to the Skyrme crystal. 

This supports the idea that a transition of some kind must take place within this generalized model, between the crystalline phase of the standard Skyrme model and the perfect fluid phase of the BPS model. A quantitative prediction of the pressure value $p_{PT}$ where this transition occurs, as well as the determination of its character---a smooth crossover or a phase transition---would require the knowledge of the full solution or, at least, the value of the parameter $\lambda$, because the contribution to the energy per baryon of the sextic term strongly depends on $\lambda$.

In \cite{Adam_2015a}, the BPS submodel was used to model the full neutron star core and, therefore, the model parameters $\lambda$ and $\mu$ were fitted to match with the infinite nuclear matter approximation at zero pressure. In the present case, however, the Skyrme crystal describes the low-pressure region and, therefore, should be fitted to nuclear matter. 
In this section, we will propose an EoS for the generalized model corresponding to $\lag_{SK}^{gen}$. The value of $\lambda$ will be determined, instead, by the behavior of the EoS in the limit of very high pressure, in which, as argued, it can be approximated by only the sextic term, see below.

\begin{figure}[b!]
\hspace*{-0.4cm}
    \centering
    \includegraphics[scale=0.4]{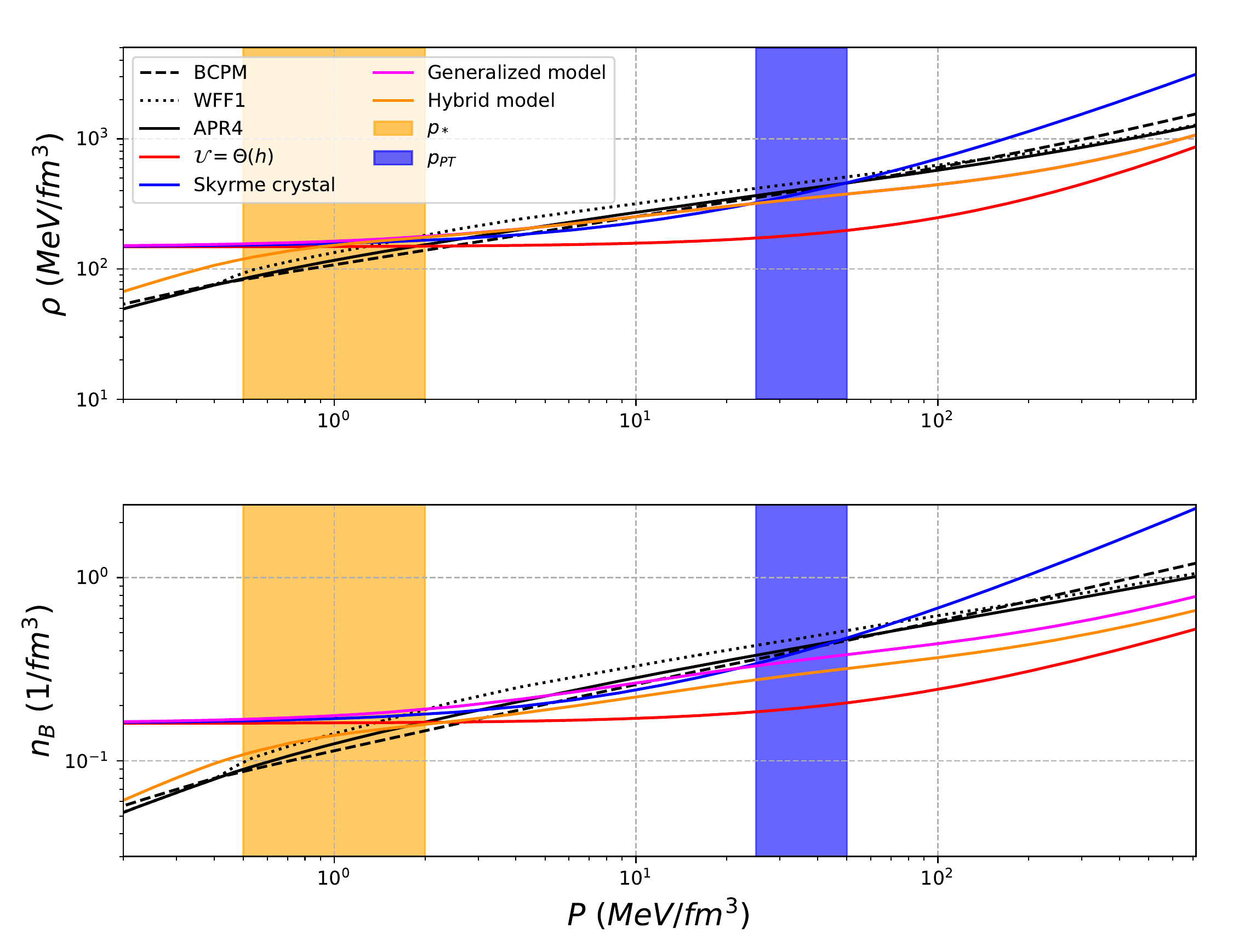}
    \caption{Comparison of the Skyrme crystal and the generalized model EoS to other neutron star  EoS usually considered in the literature, namely, APR4 \cite{Akmal:1998cf}, WFF1 \cite{Wiringa:1988tp} and BCPM \cite{Vinhas_2015} (the BCPM EoS turns out to be numerically very similar to SLy4 \cite{Douchin:2001sv}). In this figure, we show the curve for  the generalized model with $p_{PT}=50$ MeV/fm$^3$, but the range of possible values of $p_{PT}$ ($p_{*}$) that yield results consistent with observations corresponds to the blue (yellow) stripe: $p_{PT} \in [25, 50]$ MeV/fm$^3$, $p_* \in [0.5, 2.0]$ MeV/fm$^3$.}
    \label{fig:EoS}
\end{figure}

From the previous considerations, we can construct a generalized EoS which takes into account both the standard Skyrme and BPS submodels at different regimes, based on simple assumptions on the behavior of the full solutions  in the low and high pressure regimes, without knowing these solutions explicitly. Indeed, we will assume that the low pressure solutions of the complete model are still Skyrme crystals whose energy is approximately described by \eqref{energy/baryon}.  
In the fluid high-pressure phase, we will assume that the sextic term provides the most important contribution, and the complete solutions can be well described by a BPS Skyrme model. We can model this behavior by introducing a certain value of the pressure, $p_{PT}$, above which the solutions are described by a BPS fluid. Therefore, the generalized EoS $\rho_{\text{Gen}}(p)$ must satisfy
\begin{equation}
    \rho_{\text{Gen}}(p)\simeq\left\{\mqty{\rho_{SK}(p),&\quad p<< p_{PT}\\\text{const.}+p,&\quad p>> p_{PT}.} \right.
    \label{limitsGenEoS}
\end{equation}
A simple way of parametrizing this behavior that yields a smooth transition between these two regimes is to consider an EoS of the form
\begin{equation}
\rho_{\text{Gen}}(p)=(1-\alpha(p))\rho_{SK}+\alpha(p)(p+\rho_{SK}(p_{PT})),
\label{GenEoS}  
\end{equation}
where $\alpha (p)$ is a function which interpolates between the two regimes, i.e., 
$\alpha \to 0$ for $p/p_{\rm PT} \to 0$ and $\alpha \to 1$ for $p/p_{\rm PT} \to \infty$. Concretely, we consider the interpolating functions
\begin{equation}
    \alpha(p,p_{\rm PT},\beta)=\frac{\left( \frac{p}{p_{PT}}\right)^\beta}{1+\left( \frac{p}{p_{PT}}\right)^\beta}
    \label{interpol}
\end{equation}
as in \cite{Adam:2020djl}. Here, smaller values of $\beta$ produce a more gradual transition, whereas larger values correspond to a faster transition between the two regimes. For the transition between the Skyrme crystal and the BPS fluid at $p_{\rm PT}$, we have to choose the rather gradual transition $\beta = 0.9$, because otherwise the resulting energy density \eqref{GenEoS} would lead to acausal propagation (a speed of sound larger than one) in some regions inside the star for some values of $p_{PT}$.

As a result of this interpolation, the energy density contribution from the crystal becomes less and less important as $p$ grows, freezing at its value at $p_{PT}$ for sufficiently high pressures, playing the role of an effective potential energy for the BPS Skyrme model. The $p$ dependence for $p>p_{PT}$ is taken into account by $\rho_6$, which is known to provide the leading contribution for large $p$. Therefore, the generalized EoS \eqref{GenEoS} is effectively equivalent to that of a BPS Skyrme model with a theta potential \cite{Adam:2013wya} for $p>>p_{PT}$. In the following section, we will see that the value of $p_{PT}$ determines the maximum mass of a NS, so we may adjust the value of $p_{PT}$ to agree with the current maximum mass limit for NS.

To obtain the baryon density $n$ in the generalized model, we use the well-known Euler relation
 $   {\rho = -p + \pdv{\rho}{n}\: n},$
which yields a differential equation for $n$, that we integrate using $n(p=0) \equiv n_0 = 0.16$ $\text{fm}^{-3}$ as initial condition to obtain the curve $n(p)$. The result, and the corresponding EoS $\rho (p)$, are shown in Fig. \ref{fig:EoS}, where other EoS have been included for comparison.

%%%%%%%%%%%%%%%%
\subsection{Addition of crust: the hybrid EoS}
%%%%%%%%%%%%%%%%%
The generalized Skyrme EoS (\ref{GenEoS}), by construction, only describes nuclear matter above nuclear saturation \footnote{we use the recent value  $n_0 = 0.160\:{\rm fm}^{-3}$ for the nuclear saturation density, see \cite{Vinhas_2015}}.
Below saturation density, nuclear matter in a NS is known to be in a rather inhomogeneous state, resulting from a competition between nuclear and electromagnetic forces (e.g., "nuclear pasta" phases \cite{Chamel_2008}).
In principle, the (generalized) Skyrme model can be coupled to the electromagnetic interaction, so these low-density phases are fundamentally within its scope. Full field-theoretical calculations for this coupled system and for large $\mathcal{B}$ are, however, not feasible, and a macroscopic (hydrodynamical) treatment is currently unknown. 
On the other hand, the standard methods of nuclear physics, such as many-body techniques, can be used to describe these low-density NS crust regions and are completely reliable there.  This motivates us to consider a hybrid version of \eqref{GenEoS} in which, at a sufficiently low density $n_{*}$ (or, equivalently, $p_*$), a neutron star crust EoS  $\rho_{\text{BCPM}}(p)$ is glued, 
\begin{equation}
    \rho_{\text{Hyb}}(p)\simeq\left\{\mqty{\rho_{\text{BCPM}}(p),&\quad p\leq p_{*}\\\rho_{\text{Gen}}(p),&\quad p\geq p_{*}.} \right.
    \label{HybridEoS}
\end{equation}
Concretely, we choose the BCPM EoS of \cite{Vinhas_2015}, based on the Brueckner-Hartree-Fock (BHF) approach (plus the
BCPM density functional for the crust). For the crust and the outer core $n\lesssim n_0$, nuclear matter is well understood, and standard nuclear physics EoS like \cite{Vinhas_2015} should provide a precise description of NS matter. Again, we choose a smooth transition between the two regimes, using the interpolating function (\ref{interpol}). 
Now we choose the faster transition $\beta=2$, exactly as was done in \cite{Adam:2020djl} (replacing $p_{\rm PT}$ by $p_*$). 

%%%%%%%%%%%%
\section{Observational constraints}
%%%%%%%%%%%%%
To determine the static properties of the resulting NS, we simply insert the hybrid EoS \eqref{HybridEoS} into the relativistic equations of hydrodynamical equilibrium, the so-called Tolman-Oppenheimer-Volkoff (TOV) equations  \cite{Tolman:1939jz,Oppenheimer:1939ne}.
In this hybrid EoS, there are only two free parameters, namely the values of $p_{*}$ and $p_{PT}$ corresponding to the low and high density parts of the hybrid EoS. Here we show that recent astrophysical and gravitational wave observations actually tightly constrain the value ranges for both parameters.
 For example, from the mass-radius curves for different values of these parameters, we find that only the value of $p_{PT}$ affects the maximum NS mass in the model. Thus, we could for example constrain the value of $p_{PT}$ using the maximum mass limit for nonrotating NS of $M/M_{\odot}=2.16^{+0.17}_{-0.15}$ proposed in \cite{Rezzolla_2018}. However, given the recent GW observations of GW190425, with a total mass of $3.4^{+0.3}_{-0.1}M_{\odot}$ and mass ranges of components varying from 1.12 to 2.52 $M_\odot$ \cite{Abbott:2020uma} and GW190814, a compact binary merger between a $22.2-24.3 M_{\odot}$ black hole and a secondary object which falls in the mass gap  ${(2.50-2.67 M_{\odot})}$ \cite{Abbott:2020khf}, we have allowed the range of values of $p_{PT}$ to yield stars of maximum mass up to $\sim 2.7\, M_{\odot}$
 \begin{figure}
     %\hspace*{-0.4cm}
     \includegraphics[scale=0.55]{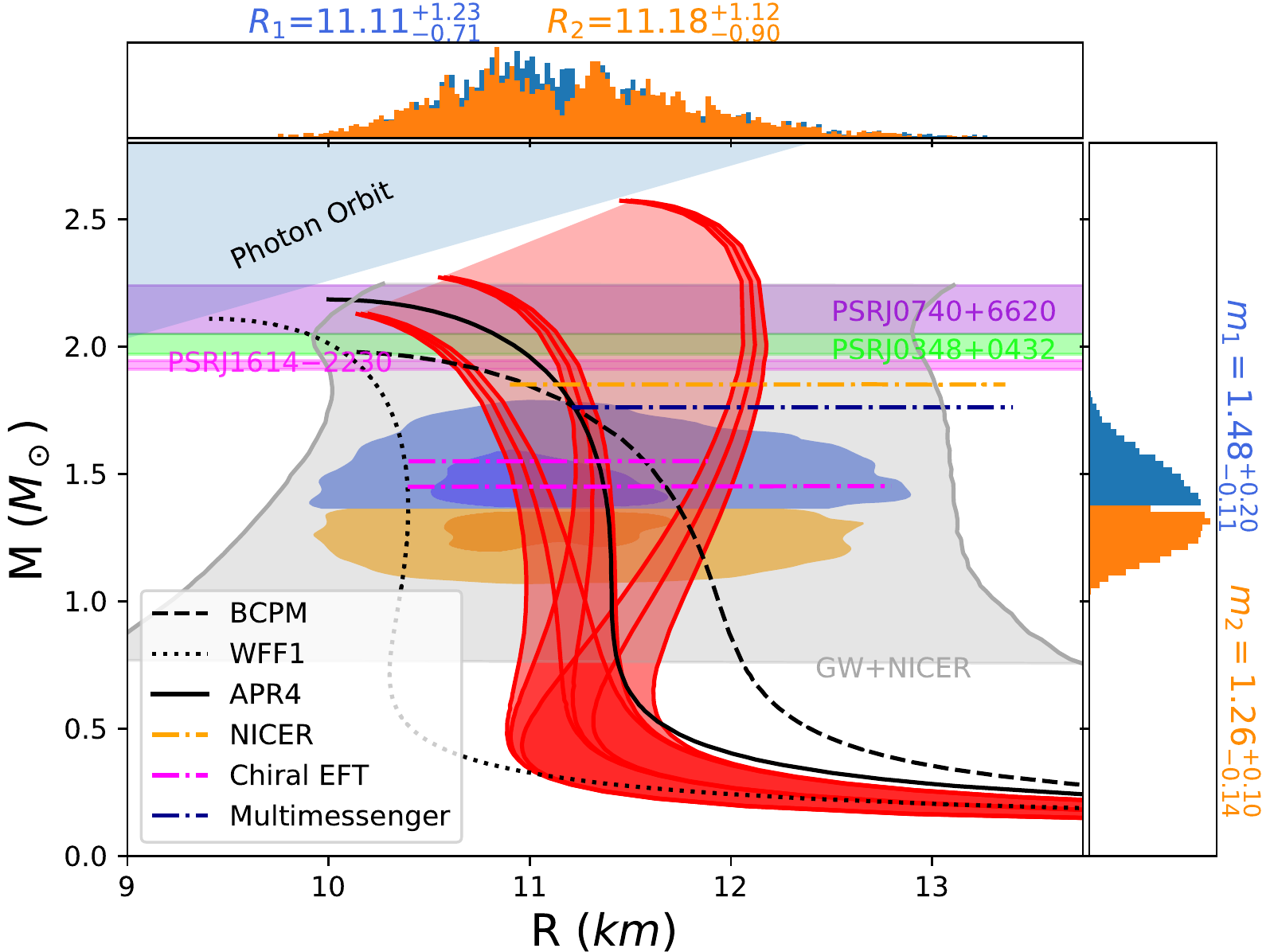}
     \caption{Mass-Radius relation for the hybrid model (red curves) for different combinations of values of ${p_{*}=\{0.5,1,2\}\text{MeV/fm}^{3}}$ and ${p_{\text{PT}}=\{25,40,50\}\text{MeV/fm}^{3}}$. The red shaded region corresponds to the accessible region of the hybrid model with $p_{*}$ and $p_{PT}$ within the given ranges (see Fig. 1).}
     \label{fig:MvR}
 \end{figure}

 In Fig. \ref{fig:MvR} we show different mass-radius curves of the hybrid model corresponding to different values of $p_{PT}$. We can see a good agreement, for any pair $(p_{*},p_{\text{PT}})$ within the ranges ${p_*\in[0.5,2]\,\text{MeV/fm}^{3}}$ and ${p_{PT}\in[25,50]\,\text{MeV/fm}^{3}}$, with the most likely mass-radius relation for the NS corresponding to the GW170817 event \cite{Abbott_2017}. We haven't included the corresponding data of the second BNS event, namely, GW190425, since it was less informative on matter effects than GW170817 , although our data is still compatible with this event as well, specially for lower values of $p_{PT}$  \cite{Abbott:2020uma}. In the same figure, we represent the masses of some of the heavier pulsars measured by the NICER collaboration, PSR $\mathrm{J} 1614-2230\left(1.928 \pm 0.017 \mathrm{M}_{\odot}\right)$\cite{Fonseca:2016tux}, PSR {$\mathrm{J} 0348+0432$}$\left(2.01 \pm 0.04 M_{\odot}\right)$ \cite{Antoniadis:2013pzd} and {$\mathrm{PSR}\,\mathrm{J}0740+6620$} $\left(2.14_{-0.09}^{+0.10} M_{\odot}\right)$ \cite{Cromartie:2019kug}, as well as the most probable M-R region from combined observations of GW and these heavy pulsars \cite{Landry:2020vaw}. Also, other constraints from NICER, chiral EFT and multimessenger observations are represented, adapted from \cite{Constraints2020neutron} and \cite{Greif:2020pju}. 
 
The observed gravitational waveform can also be used to
place direct constraints on the tidal deformability of NS. Indeed, the  waveform produced by the coalescence of two NS at the early phase of the inspiral depends on the underlying EoS mostly
through the tidal Love number \cite{Hinderer:2009ca}. However, the individual Love numbers for the two stars
cannot be disentangled in the observed gravitational waveform.
Instead, what is measured is the so-called effective tidal deformability $\tilde{\Lambda}$, a mass weighted average of the deformabilities of the individual stars in the merger \cite{Flanagan_2008}. Similarly, the two component masses are not measured directly, but the chirp mass, $M_c=m_{1} \,{q^{3 / 5}}/{(1+q)^{1 / 5}}$
where $q=m_1/m_2$ is the mass ratio,
can actually be tightly constrained. In the case of the GW170817 event, the chirp mass was constrained to $1.188_{-0.002}^{+0.004}$
at the $90\%$ confidence level, and  the mass ratio was constrained to be in the range $0.7-1$ within the same confidence level, whereas the effective tidal deformability was inferred to be smaller than $800$ \cite{Abbott:2018wiz}.

\begin{figure}[bht]
\hspace*{-0.5cm}
    \includegraphics[scale=0.4]{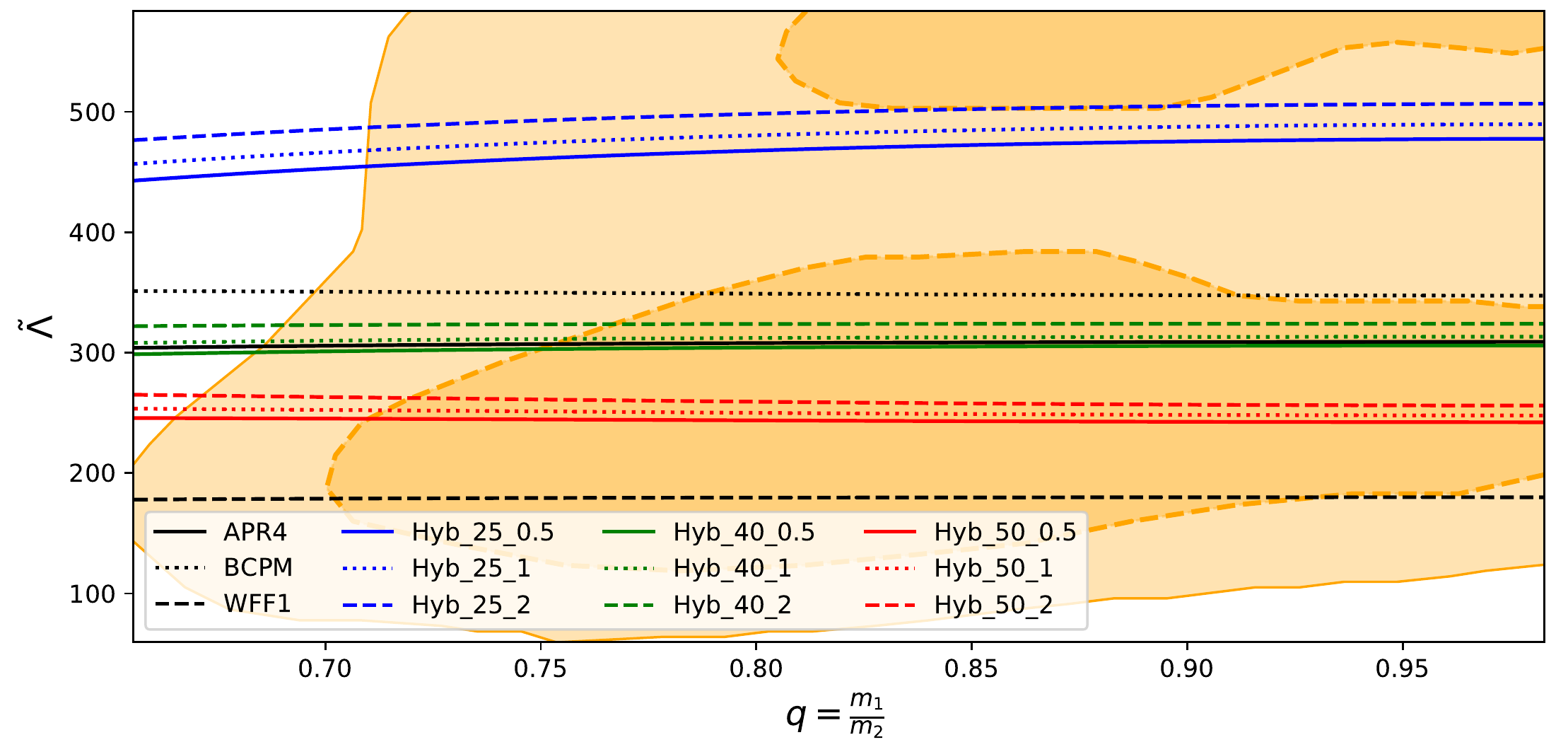}
    \caption{$\tilde \Lambda$ as a function of the mass ratio. The orange shaded regions correspond to the $50\%$ (dark) and $90 \%$ (light) credible regions for the joint posterior of $\tilde\Lambda$ and $q$ PDFs as obtained in \cite{Abbott:2018wiz} assuming a low spin prior. Notation for curves from the EoS (\ref{HybridEoS}): Hyb\_\_$p_{PT}$\_\_$p_*$.
    }
    \label{fig:LAMBDATIDAL}
\end{figure}

Such measurements allow to reduce the set of Skyrme models able to reproduce the NS properties. Following \cite{Yagi_2013}, we have solved the Einstein equations for slowly rotating Skyrmion stars with the hybrid EoS using the Hartle-Thorne formalism \cite{HartleI,HartleThorne85} and obtained the dimensionless tidal deformability of stars described by this model as a function of their TOV mass. On the other hand, since the chirp mass
of the binary progenitor of GW170817 is well measured,
for any given EoS the effective deformability reduces to a simple EoS-dependent function of the mass ratio.
These curves, together with the constraints commented above, are represented in Fig. \ref{fig:LAMBDATIDAL}, from where it follows that our new EoS is compatible with the data from \cite{Abbott:2018wiz} for the ranges of $p_{*}$ and $p_{\text{PT}}$ considered. Future measurements of the tidal deformability of NS will allow us to further constrain these ranges, since we find that the curves $\Tilde{\Lambda}(q)$ depend on the particular values of both parameters.

%%%%%%%%%%%%%%%%
\subsection{The sextic term and the $\omega$ meson}
%%%%%%%%%%%%%%%%%
As explained, the sextic term \eqref{sextic} automatically provides the leading contribution at high densities if it is included in the effective action. On the other hand, this term is physically justified because it provides the leading contribution of the $\omega$ meson repulsion in a derivative expansion which results from integrating out the $\omega$ field from an extended Lagrangian which includes both pions and vector mesons \cite{ADKINS1984251,Meissner:1987ge,Vecmes2}.  This relation to the $\omega$ meson not only motivates the sextic term, but it also leads to an expression of its coupling constant $\lambda$ in terms of the physical parameters of the $\omega$ meson. Indeed, it can be shown \cite{InmediumBPS} that 
$ \lambda^2=\frac{g_{\omega}^2}{2\pi^4m_\omega^2}$,
 where $m_\omega$ and $g_\omega$ are, respectively, the mass and coupling constant of the $\omega$ vector meson. For the empirical values $m_\omega = 783 \,\text{MeV}$ and
 $g^2_\omega/(4\pi)\sim 12$ \cite{Meissner:1987ge}, we get $\lambda^2 \sim 10\, \text{MeV\,fm}^3$.

The $\lambda$ coupling constant appearing in front of the sextic term in the generalized Lagrangian does not directly show up in the generalized EoS proposed in this work, whose parameters are constrained by the observations of maximum mass and deformability. However, the generalized Skyrme model EoS approaches the EoS of the BPS submodel for sufficiently large pressure, by assumption. We can, therefore, extract an effective value of $\lambda$ by taking the limit of infinite pressure and using the BPS EoS \eqref{BPSEoS}, which immediately implies
$
    n^2=(\rho+p)/{(2\lambda^2\pi^4)}.
$
The effective value of $\lambda$ is then given by
\begin{equation}
    \lambda_{eff}=\lim\limits_{p\rightarrow\infty}\frac{1}{n(p)\pi^2}\sqrt{\frac{\rho_{Gen}(p)+p}{2}}.
\end{equation}
For the range of values $p_{PT}\in[25,50]\,\text{MeV/fm}^{3}$, we find that $\lambda_{eff}^2\in[10 ,13.5]\,\text{MeV\,fm}^{3}$. The values so obtained for $\lambda_{eff}^2$ are, therefore, perfectly compatible with the values obtained by assuming that the sextic term in the generalized Lagrangian results from integrating out the $\omega$ vector meson. 

\section{Conclusions}
In this letter, we propose a completion of standard nuclear physics EoS at low densities---known to be reliable there---by an EoS based on the generalized Skyrme model in the uncharted territory above nuclear saturation density $n_0$. In the simplest version of Skyrme models, where electromagnetic effects, quantum corrections or the proton-neutron mass difference are not taken into account, they can describe nuclear matter only for $n\ge n_0$, by construction. The use of the generalized Skyrme model at densities $n_0<n\leq n_{\text{max}}$ is based on the assumptions that $i)$ strong-interaction effects (nuclear repulsion) are more important than degeneracy pressures in that region, $ii)$ the extended character of nucleons---which is automatic in the Skyrme model---is relevant at high pressure and $iii)$ nucleons are the only relevant DoF inside NS cores (no exotic contributions). This last assumption is shared by many NS models. Here, $n_{\text{max}}$ corresponds to the central density of the maximum mass NS, which is $n_{\text{max}}=7.1 n_0$ for $p_{PT}=50{\rm MeV}/{\rm fm}^3$ and $n_{\text{max}}=5.3 n_0$ for $p_{PT}=25{\rm MeV}/{\rm fm}^3$, safely below the deconfinement phase transition density $\sim 40n_0$.\footnote{In the recent paper \cite{cite-NaturepQCD}, some arguments in favor of quark matter cores of heavy NS were given. Their analysis, however, is based on the assumption that the speed of sound does not exceed the so called ``conformal bound'' $c^2_s=1/3$, which obviously does not hold in our model.  }

We find that the resulting EoS provides an excellent description of NS properties, compatible with all constraints, among them the latest ones from LIGO. Our EoS contains two parameters which have a clear physical interpretation as transitions between standard nuclear matter and the Skyrme crystal ($p_*$) and between this crystal and a Skyrme fluid ($p_{PT}$). In particular, we propose a rather smooth transition between a crystalline and a fluid regime for $20\le p_{PT}\cdot{\rm fm}^3/{\rm MeV}\le 50$, whose precise position may be determined by more precise NS binary observations. Let us remark that the very recent observation of the GW190814 event \cite{Abbott:2020khf}, with a certain indication of an NS with a mass of about 2.6 $M_\odot$ \cite{fattoyev2020gw190814,Huang:2020cab}, can be easily accommodated by our generalized Skyrme model EoS, by simply choosing a slightly lower value of $p_{\rm PT}\sim 25$ MeV fm$^{-3}$ for the transition between Skyrme crystal and BPS fluid, see Fig. \ref{fig:MvR}.

We also find that the range of values for the effective coupling constant $\lambda$ of the sextic term which results from our generalized EoS and the fit to realistic NS is perfectly compatible with the range of values resulting from its relation to the $\omega$ meson.

Motivated by the results obtained with the hybrid EoS proposed above, it
would certainly be interesting to try to derive a similar EoS from an
exact solution of the generalized Lagrangian, for example, using a
crystalline ansatz for the Skyrme field, as in \cite{gencryst}. One then
could study whether this exact EoS presents a phase transition of some
kind, in the same fashion as the proposed hybrid EoS. We leave this study
for a future work.

Finally, we would like to comment on the similarities and differences of
our proposal to the scenarios considered in Ref. [7]. The calculations and
discussions in [7] are based on the (standard) Skyrme crystal and are, in
this sense, similar in spirit to ours. There are, nevertheless, some
important differences. First of all, in [7] more degrees of freedom are
considered, among them the dilaton to recover the scale symmetry of QCD at
large densities. At low density, this symmetry is broken spontaneously—the
dilaton freezes—and the Skyrme model is recovered. Further, higher mass
mesons are taken into account implicitly in \cite{Ma:2019ery}. The main difference for
the present purpose, however, is related to another phase transition which
is known to occur in the standard Skyrme crystal \cite{Kugler:1989uc}, namely the
transition from a skyrmion phase to a half-skyrmion phase as density
increases. In \cite{Ma:2019ery}, this transition leads to a significant stiffening of
the EoS which is mainly related to an enhanced contribution of the
symmetry energy in the half-skyrmion phase. In our case, we assume that in
the region where the Skyrme model is effective (i.e., for $p>p_*$) we are
always in the half-skyrmion phase. In addition, the Skyrme crystal
influences the EoS only via its scaling properties in our case, see Eq.
(1), and the effects of the symmetry energy are taken into account only
implicitly, by an appropriate choice of our physical parameters. The
stiffening of the EoS is caused, instead, by the sextic term, i.e., by the
$\omega$ repulsion, as explained in the main text.

\begin{acknowledgements}
The authors acknowledge financial support from the Ministry of Education, Culture, and Sports, Spain (Grant No. FPA2017-83814-P), the Xunta de Galicia (Grant No. INCITE09.296.035PR and Conselleria de Educacion), the Spanish Consolider-Ingenio 2010 Programme CPAN (CSD2007-00042), Maria de Maetzu Unit of Excellence MDM-2016-0692, and FEDER. 
\end{acknowledgements}

% The \nocite command causes all entries in a bibliography to be printed out
% whether or not they are actually referenced in the text. This is appropriate
% for the sample file to show the different styles of references, but authors
% most likely will not want to use it.
%\bibliographystyle{unsrt} 
\bibliography{PRLbiblio}% Produces the bibliography via BibTeX.

\end{document}